\documentclass[a4paper]{jpconf}
\usepackage{graphicx}
\def\beq{\begin{equation}}
\def\eeq{\end{equation}}
\def\bea{\begin{eqnarray}}
\def\eea{\end{eqnarray}}

\usepackage[square,numbers,sort&compress]{natbib}

\begin{document}
\title{The role of pressure anisotropy in the turbulent intracluster medium}

\author{M. S. Nakwacki$^{1,}$\footnote[2]{Present address:
Instituto de Astronom\'ia y F\'isica del Espacio, UBA-CONICET, cc 67, suc 28, cp 1428, CABA, Argentina}, 
E. M. de Gouveia Dal Pino$^1$,
G. Kowal$^{1}$, and R. Santos-Lima$^1$}
\address{$^1$ Instituto de Astronom\'ia, Geof\'isica e Ci\^encias Atmosf\'ericas, Universidade de S\~ao Paulo, 
Rua do Mat\~ao 1226, Cidade Universit\'aria, 05508-090 S\~ao Paulo, 
Brazil}

\ead{sole@astro.iag.usp.br}

\begin{abstract}
In low-density plasma environments, such as the intracluster medium (ICM), the Larmour frequency is
much larger than the ion-ion collision frequency.
In such a case, the thermal pressure becomes anisotropic with respect to
the magnetic field orientation and the evolution of the turbulent gas is more
correctly described by a kinetic approach. 
A possible description of these collisionless scenarios is given by the so-called kinetic magnetohydrodynamic
(KMHD) formalism, in which particles freely stream along the field lines, while
moving with the field lines in the perpendicular direction. In this way a fluid-like
behavior in the perpendicular plane is restored. 
In this work, we study fast growing magnetic fluctuations in the smallest scales 
which operate in the collisionless  plasma that fills the ICM. In particular, we focus on the impact of a particular evolution of the pressure anisotropy and its implications for the turbulent dynamics of observables under the conditions prevailing in the ICM. We present results from numerical simulations and compare the results which those obtained using an MHD formalism.

\end{abstract}

\section{Introduction}
\vspace{0.2cm}
Magnetic fields play an important role in the development of large-scale
structure in the Universe, and in recent years their presence in galaxy clusters has been unambiguously
proved. According to the standard scenario of structure formation, galaxy clusters are built-up by gravitational
merger of smaller units, such as groups and sub-clusters. They are composed of hundreds of 
galaxies in a Mpc-size region, and the Intra Cluster Medium (ICM) is filled with hot and rarefied
gas, emitting in the soft-X ray domain through optically thin bremsstrahlung, magnetic fields and
relativistic particles. Magnetic fields in the ICM are investigated through synchrotron emission of
cluster-wide radio sources and from the study of the Rotation Measure of radio galaxies revealing the turbulent nature of these 
fields \citep{bonafede10a}.

Moreover, the ICM is weakly magnetized and nearly collisionless, i.e., the gyro-frequency
of both the electrons and the ions is much greater than the collision frequency (under these conditions the mean free path of ions is $\approx 30$ kpc in the hot ICM, \cite{schekochihin06}). Plasmas
with such characteristics are known to present anisotropic pressures with respect to the
magnetic field orientation \citep[see e.g.][]{Barakat82,krall,quest96}, whose imprints can survive for considerably long periods compared with
the dynamical timescales of the system. The presence of mergers, accretion, active galactic
nuclei (AGN), galactic winds and instabilities cause complex flows that generate shocks,
discontinuities and shear. All these processes lead to the generation of turbulence, which
transports and amplifies the magnetic fields present \citep{nakwacki11}.

Understanding the role and evolution of magnetic fields
in clusters of galaxies is of significant importance for many
questions including the origin of cluster magnetic fields, the
interaction of AGN with the ICM, and physical processes operating within
the ICM plasma. Which
(if any) physical mechanism dominates in the ICM depends
sensitively upon the magnitude and distribution of turbulence,
which is currently only poorly understood \citep{bogdanovic10}.

The fact that the ICM is a low density environment makes the typical MHD description not reliable. 
As a consequence, these astrophysical plasmas are known to present anisotropic
pressures with respect to the magnetic field orientation, which can be generated by several different processes, such
as kinetic pressure of cosmic rays, supernovae explosions, stellar winds or anisotropic turbulent
motions \citep[see][]{quest96,kowal11}. On small scales, this turbulence is
often expected to consist of highly anisotropic fluctuations with frequencies small compared to the ion cyclotron
frequency. For a number of applications, the small scales are also collisionless, so a kinetic treatment of the
turbulence is necessary, thus making a Kinetic MHD (KMHD) description more appropriate. 

The purpose of this contribution is to show and discuss some recent advances in the KMHD description of the ICM plasma. After a brief discussion of the properties and applicability of the Chew-Golberger-Low (CGL) closure, we will describe the KMHD framework, which does include a pressure anisotropy in the description of plasma systems. We will then discuss the impact of this pressure anisotropy on the turbulent evolution of MHD observables, using as an illustrative case a particular variation in time for the pressure anisotropy ratio that reaches the isotropic, thus MHD, state. We will show numerical results corresponding to both developments and briefly discuss implications for this scenario.  

\vspace{0.4cm}
\section{The KMHD model}
\vspace{0.2cm}
In order to determine the influence of pressure anisotropy on the turbulent evolution of
the plasma in the ICM, in this work we use
an MHD formalism with a Chew-Golberger-Low double-isothermal
closure, the so called  CGL closure \cite{cgl}, as implemented in the numerical code \citep{kowal11,kowal10,kowal07,kowal09}. As already mentioned, AGNs and the random
motion of galaxies are the main sources of kinetic energy of the ICM plasma, and may be
considered the main reason for the eventual pressure anisotropy observed in this medium.
Because the kinetic energy injection is continuous the timescale of anisotropic radiative loses
is much larger than the timescale associated with plasma instabilities, thus rendering the
CGL closure applicable. Here, we will only consider this regime since we are interested in
the large scale properties of the ICM for which this condition is satisfied.

In order to study the magnetic field dynamics in the ICM we simulate turbulence in a periodic box of $1$ Mpc of size with $128^3$ grid points solving the set of
double-isothermal KMHD equations in a conservative form given by the equations
\begin{eqnarray}
 \frac{\partial\rho}{\partial t}+\nabla . (\rho\vec{V})&=&0\\
 \frac{\partial(\rho\vec{V})}{\partial t}+\nabla . [\rho\vec{V}\vec{V}+(P+\frac{B^2}{8\pi})I-\frac{1}{4\pi}\vec{B}\vec{B}] &=&f\\
\frac{\partial\vec{B}}{\partial t}-\nabla\times(\vec{V}\times\vec{B})&=&0
\end{eqnarray}
where the term $f$ is a random solenoidal large-scale
driving force representing the turbulence driving, which is driven at wave scale $k= 2.5$, the scale injection in the model (2.5 times smaller
than the size of the box), and $P = p_\perp\hat{I} + (p_\parallel - p_\perp)\hat{b}\hat{b}$ is the pressure tensor with components $p_\parallel$ and $p_\perp$ 
parallel and perpendicular to the magnetic field direction ($\hat{b} = \vec{B}/|\vec{B}|$) \cite[see][]{kowal11}. To close the system 
a CGL closure \cite{cgl} is used, which yields:
\begin{equation}
  \frac{\partial(\rho\vec{V})}{\partial t}+\nabla .[(a_\perp^2\rho+\frac{B^2}{8\pi})I-(1-\alpha)\vec{B}\vec{B}] =f
\end{equation}
 with $p_{\perp,\parallel}=a_{\perp,\parallel}^2\rho$, where $a_{\perp,\parallel}$ are constants and represent speeds
of sound along the perpendicular and parallel directions to the magnetic field, and $\alpha=(p_\parallel-p_\perp)/(2P_{mag})$, where $P_{mag}= \frac{B^2}{8\pi}$ is the magnetic pressure.

\vspace{0.4cm}
\section{Numerical simulations}
\vspace{0.2cm}

In this work, we do not take into account 
viscosity and diffusion in the equations. The scale at which the dissipation starts to act
is defined by the numerical diffusivity of the scheme. The
numerical integration of the system evolution governed by the KMHD equations were
performed by using the second-order shock-capturing Godunov-scheme code and 
the time integration was done with the second-order Runge-Kutta method \cite{falceta08,kowal07,kowal09,kowal11}. 


The spatial coordinates are given in units of a
typical length $L_0$. The density $\rho$ is normalized by a 
reference density $\rho_0$, and the velocity field $V$ by a reference
velocity $V_0$. The constant sound speed $c_s$ is also given
in units of $V_0$ , and the magnetic field $\vec{B}$ is measured in
units of $V_0 \sqrt{4\pi \rho_0}$. Time $t$ is measured in units of $L_0/V_0$. 
 
\begin{figure}[h!]\begin{center}
\includegraphics[width=16pc]{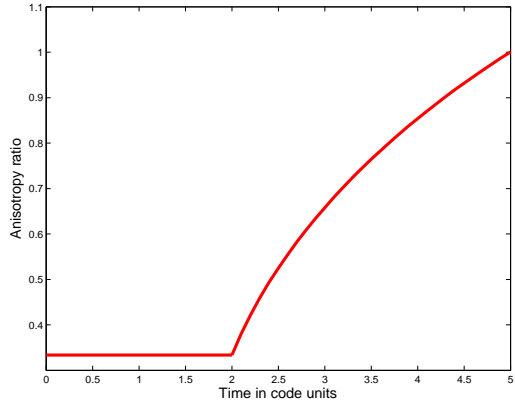}\end{center}
\caption{\label{curva}Evolution of the speed of sound ratio, $a_\parallel/a_\perp$. See the text for further details.}
\end{figure}

To go one step further in the comparison between KMHD and MHD,  
the simulation includes a variation in time for the pressure
anisotropy.  
The values for the parallel and perpendicular sound speeds
are set equal to $1.0$ and $3.0$, respectively, keeping them constant until $t = 2.0$, when the turbulence may be
considered as fully developed. From $t = 2.0$, there is a change of pressure anisotropy 
(reducing the perpendicular sound speed and increasing the parallel one) according to the following law $a_\perp(t) = a_{\perp 0} + (a_{\perp 0}-a_{\perp \infty})/(1-t_{half}/t)$,
where $a_{\parallel 0} = 1.0$ and $a_{\perp 0} = 3.0$ are the initial values of the sound speed, $a_{\perp\infty} = 1.5$ is the
perpendicular sound speed in the limit $t_\infty$, and $t_{half}$ $= 6.3$ is the time after which the
perpendicular speed decays to the value of $(a_{\perp 0} + a_{\perp\infty})/2$. As the parallel sound speed also
changes, the total square sound speed, $a_\parallel^2(t) + 2a_\perp^2(t)$, is kept constant and equal to the initial value, i.e. the total pressure is
kept constant and there is an interplay between the parallel and perpendicular pressures. In this way,
we change the pressure anisotropy in time starting from a low value ($\sim 0.3$) and ending up close to 
the isotropic value, thus reaching the MHD regime. The chosen final value of the anisotropy parameter reaches the value $\sim 0.98$ 
corresponding to the value in the ICM that is expected from theoretical considerations \cite{rosin10,Schekochihin05,howes06}. Fig. \ref{curva} shows the evolution of the anisotropy parameter.

We also consider the conditions prevailing in the ICM: superalfvenic and subsonic regime with $B_{0}\hat{x}=0.1$ as the initial magnetic field.

\vspace{0.4cm}
\subsection{Results}
\begin{figure}[h]
\begin{minipage}{12pc}
\includegraphics[width=12pc]{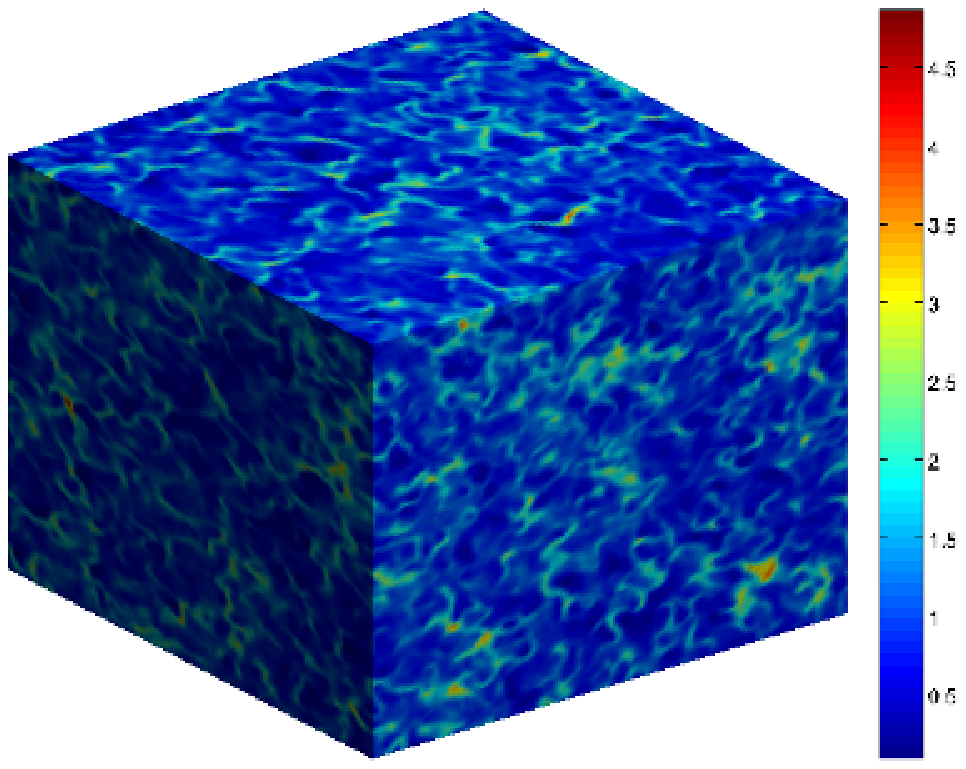}
\caption{\label{densk1}Central slices (projected on the walls of the computational domain) showing the plasma density at t=2 for KMHD.}
\end{minipage}\hspace{2pc}%
\begin{minipage}{12pc}
\includegraphics[width=12pc]{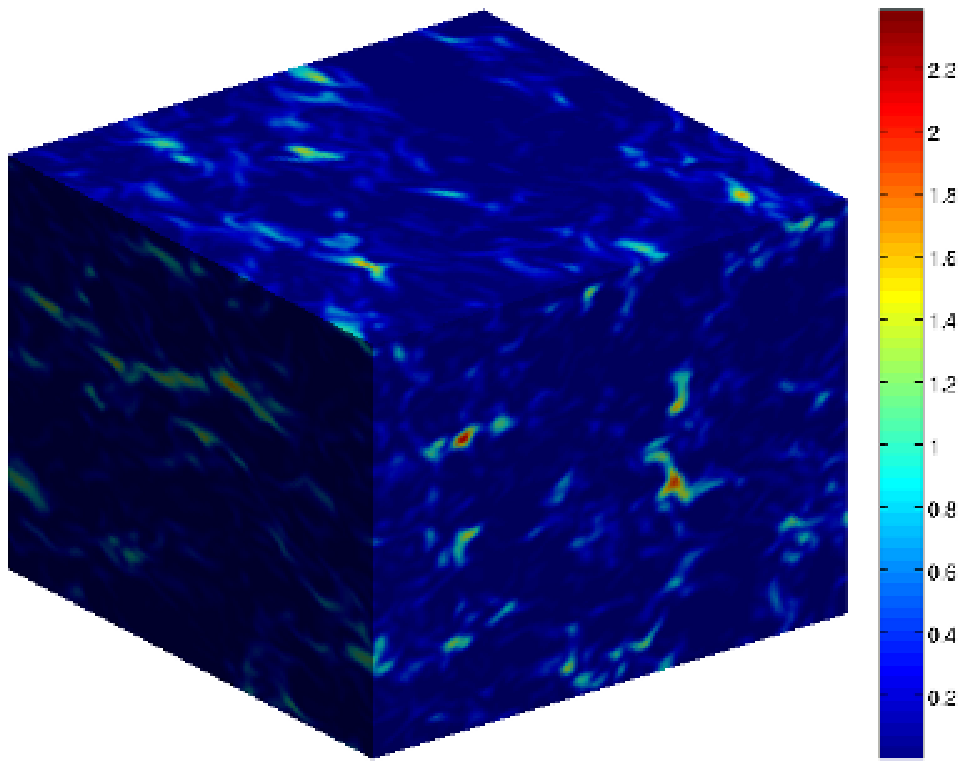}
\caption{\label{bk1}Central slices (projected on the walls of the computational domain) showing magnetic field intensity at t=2 for KMHD.}
\end{minipage} \hspace{2pc}
\begin{minipage}{12pc}
\includegraphics[width=12pc]{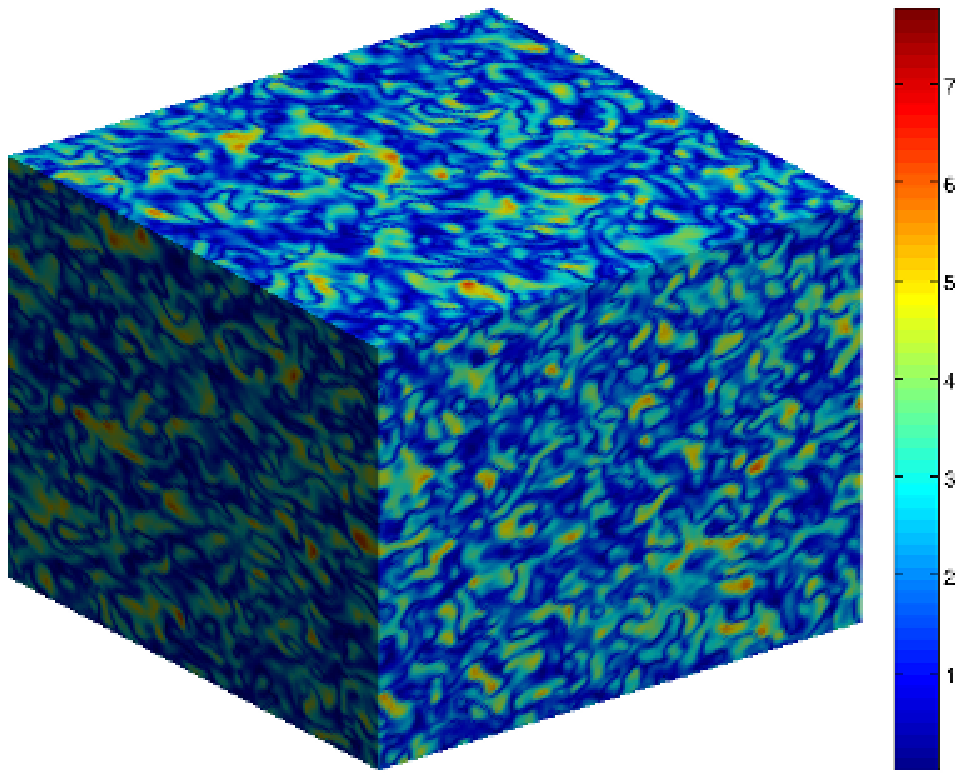}
\caption{\label{vk1}Central slices (projected on the walls of the computational domain) showing the velocity intensity at t=2 for KMHD.}
\end{minipage} 
\begin{minipage}{12pc}
\includegraphics[width=12pc]{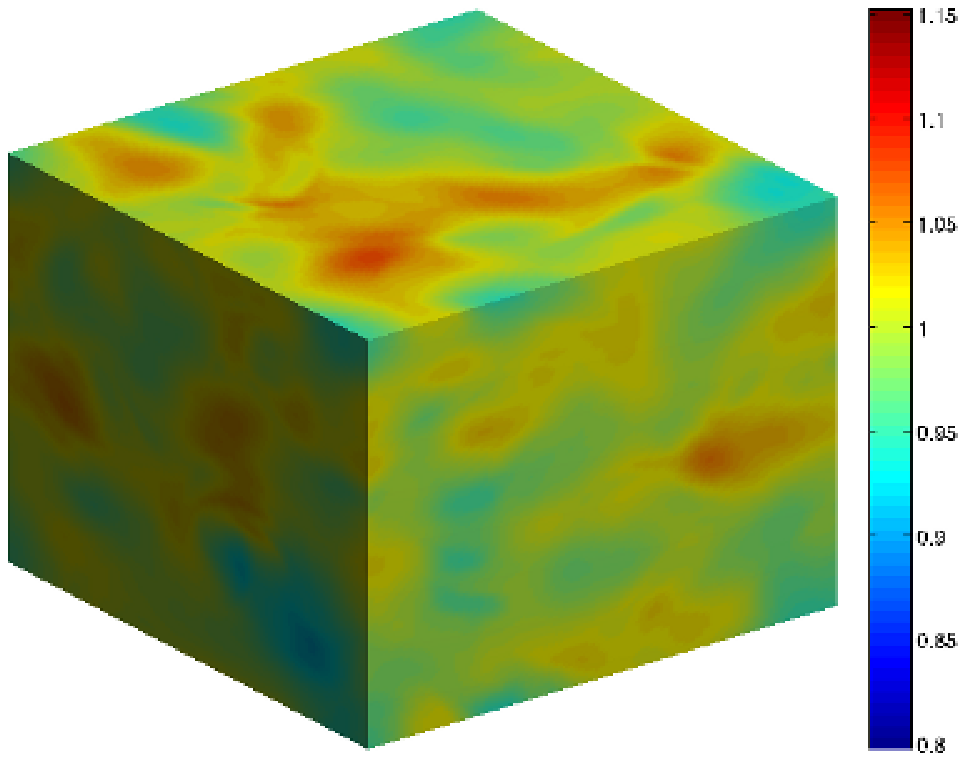}
\caption{\label{densm1}Idem Fig. \ref{densk1} for MHD.}
\end{minipage}\hspace{2pc}%
\begin{minipage}{12pc}
\includegraphics[width=12pc]{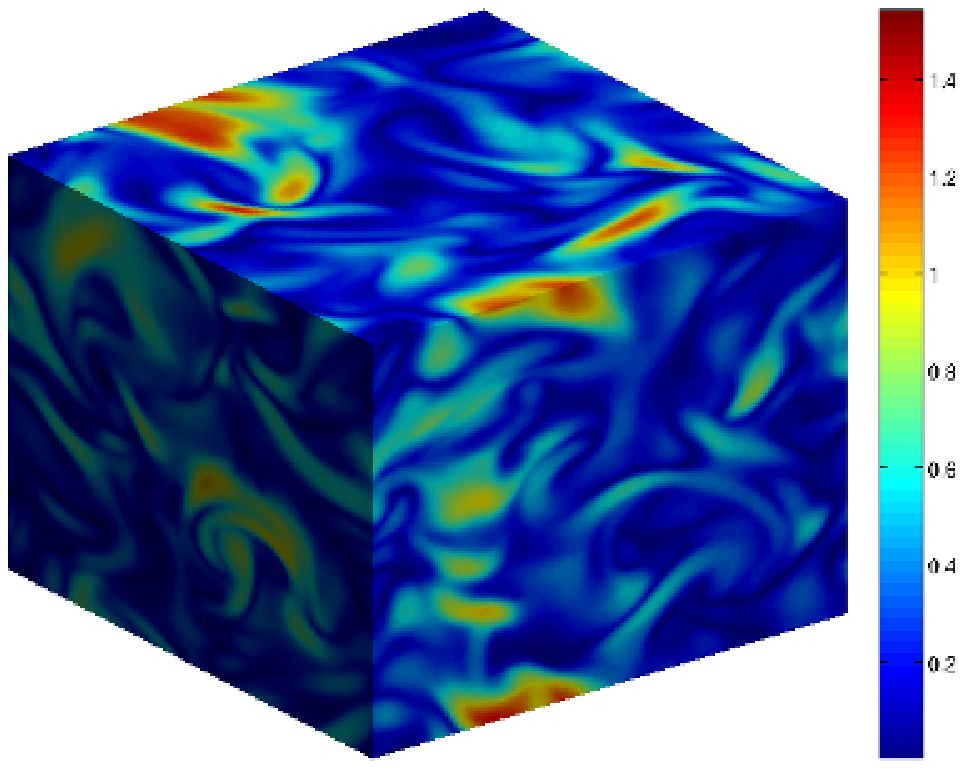}
\caption{\label{bm1}Idem Fig. \ref{bk1} for MHD.}
\end{minipage} \hspace{2pc}
\begin{minipage}{12pc}
\includegraphics[width=12pc]{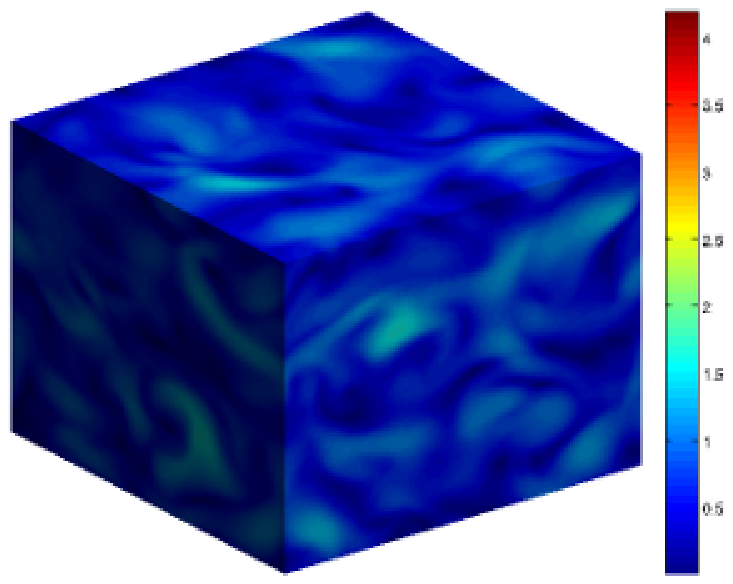}
\caption{\label{vm1}Idem Fig. \ref{vk1} for MHD.}
\end{minipage} 
\end{figure}

\vspace{0.2cm}
The results obtained for the evolution of the physical observables, namely velocity, magnetic field and density show strong differences between
the KMHD and standard MHD models. It is interesting to note that kinetic instabilities play a role in the
evolution of turbulence under the particular conditions taken into account. In Figs. \ref{densk1}-\ref{vm1}, we present the density, velocity intensity and magnetic field intensity in the central slices of the computational domain 
obtained for the KMHD model (first line), as as well as for the MHD model (second line) for comparison at t=2 when the turbulence can be considered as fully developed and the pressure anisotropy rate starts changing. It is seen that for the three quantities that we consider the structures obtained in KMHD are significantly smaller than the ones obtained in MHD. Moreover, density, magnetic field intensity and velocity fluctuations are slightly smaller in the isotropic case. 

\begin{figure}[h]
\begin{minipage}{12pc}
\includegraphics[width=12pc]{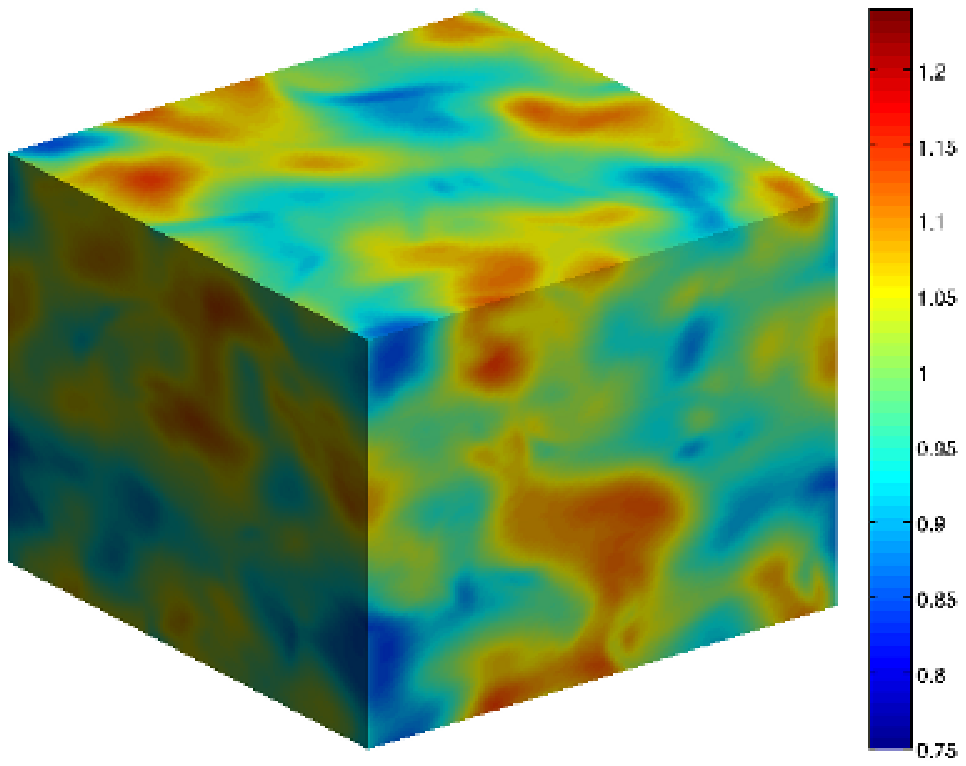}
\caption{\label{densk2}Idem Fig. \ref{densk1} at t=5.}
\end{minipage}\hspace{2pc}%
\begin{minipage}{12pc}
\includegraphics[width=12pc]{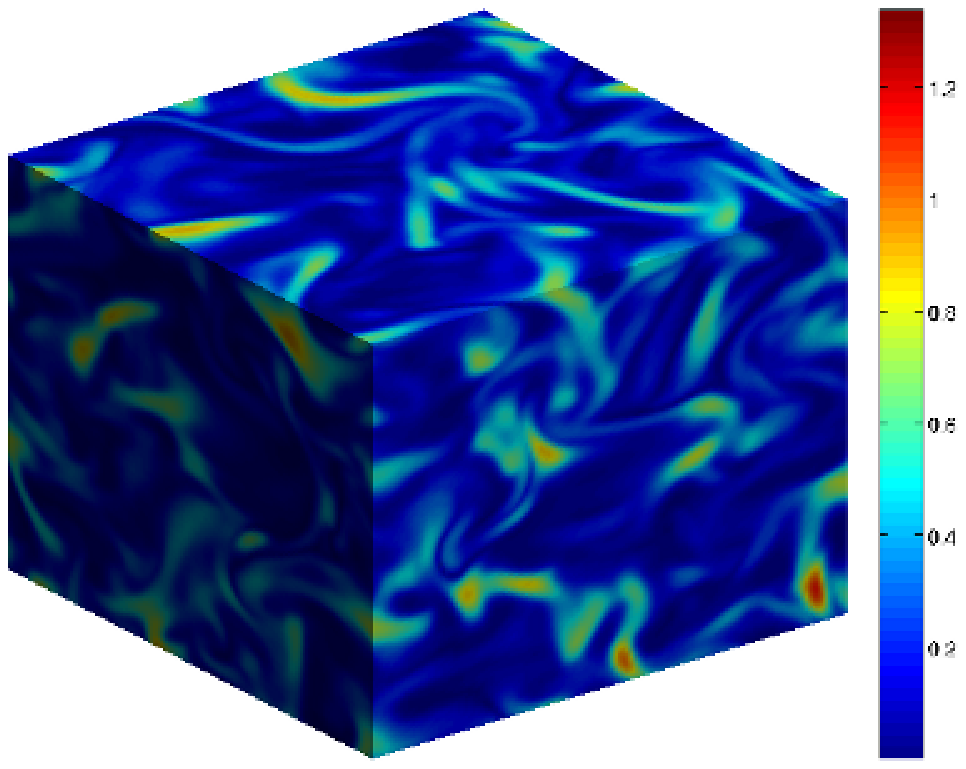}
\caption{\label{bk2}Idem Fig. \ref{bk1} at t=5.}
\end{minipage} \hspace{2pc}
\begin{minipage}{12pc}
\includegraphics[width=12pc]{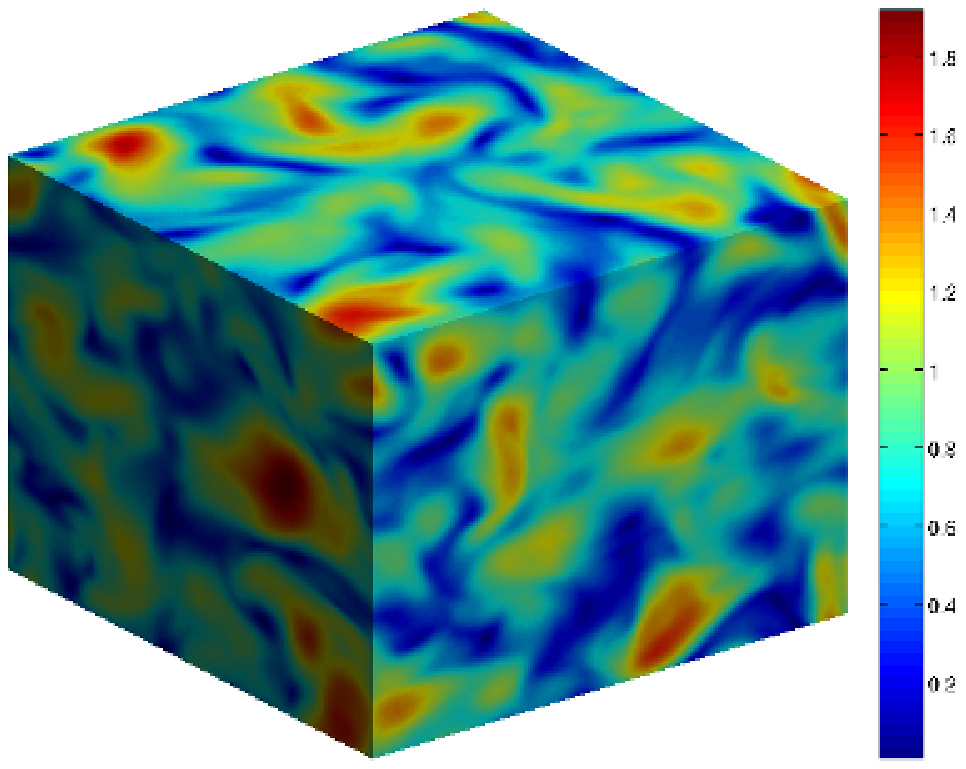}
\caption{\label{vk2}Idem Fig. \ref{vk1} at t=5.}
\end{minipage} 
\begin{minipage}{12pc}
\includegraphics[width=12pc]{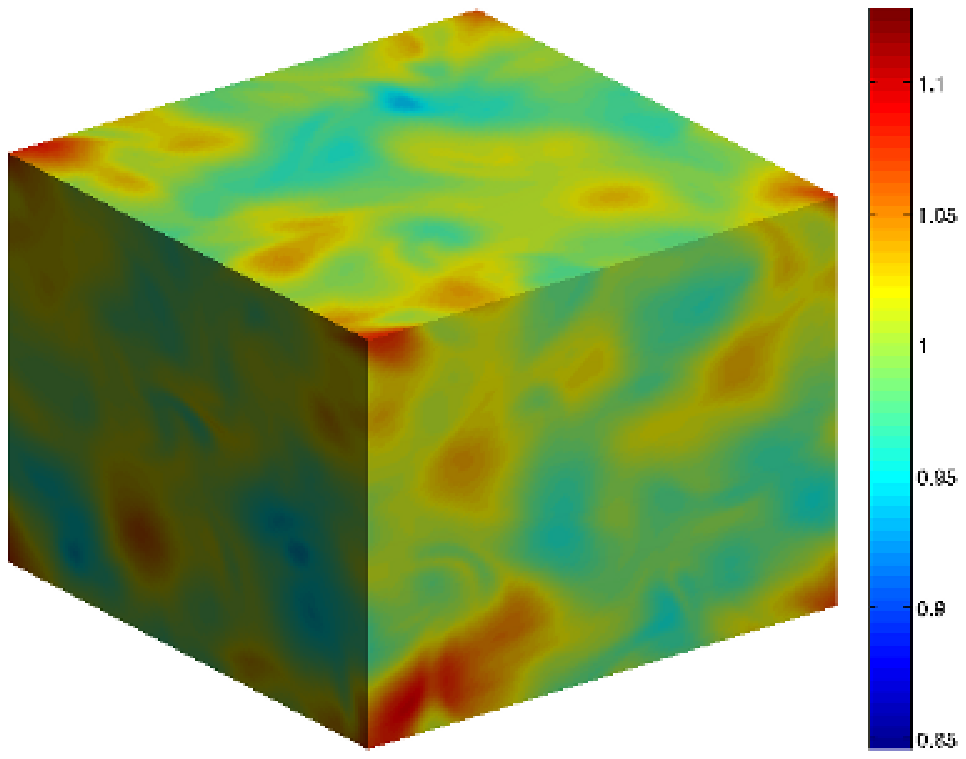}
\caption{\label{densm2}Idem Fig. \ref{densm1} at t=5.}
\end{minipage}\hspace{2pc}%
\begin{minipage}{12pc}
\includegraphics[width=12pc]{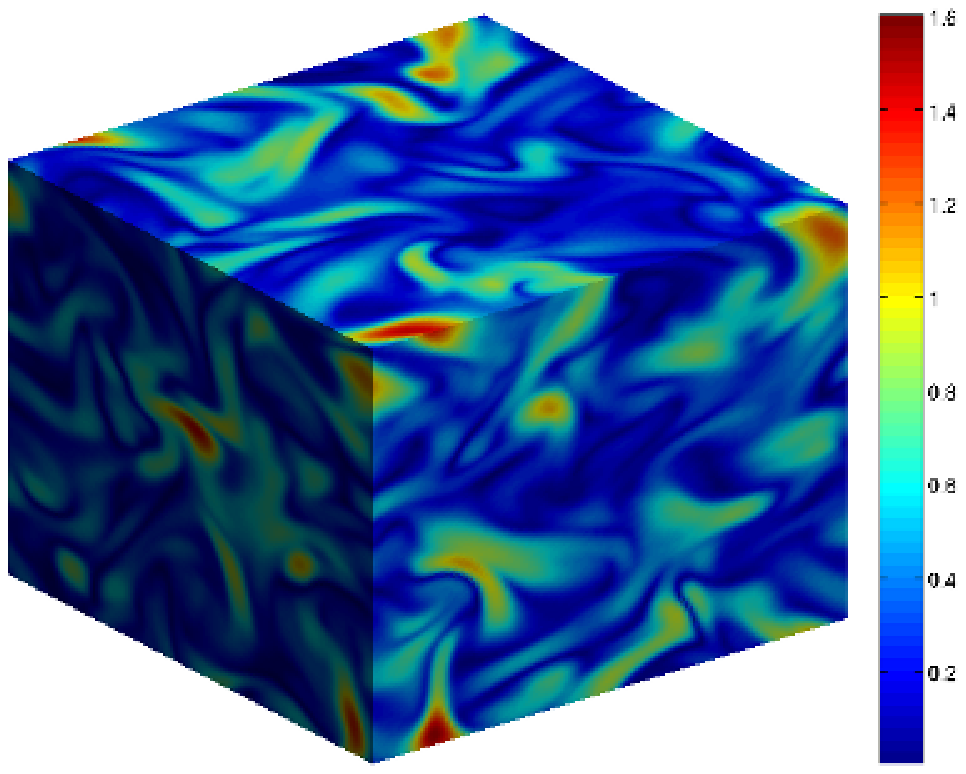}
\caption{\label{bm2}Idem Fig. \ref{bm1} at t=5.}
\end{minipage} \hspace{2pc}
\begin{minipage}{12pc}
\includegraphics[width=12pc]{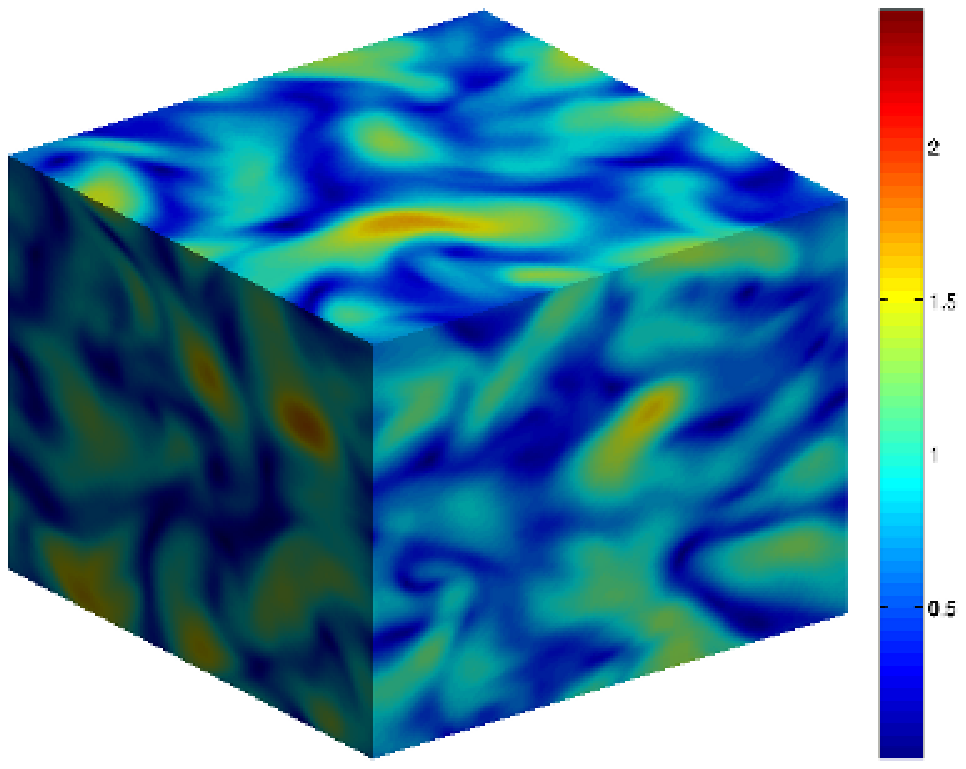}
\caption{\label{vm2}Idem Fig. \ref{vm1} at t=5.}
\end{minipage} 
\end{figure}

It is interesting to quantify the impact of the pressure anisotropy change on the evolution of the three observables mentioned above. To this end, we show in Figs. \ref{densk2}-\ref{vm2} density, velocity intensity and magnetic field intensity in the central slices of the computational domain 
obtained for the KMHD model (first line), as as well as for the MHD model (second line) at t=5. Comparing the results obtained at both times it is seen that the size of the structures in KMHD evolved towards their MHD counterpart. This is reasonable because at t=5 the pressure becomes almost isotropic in the KMHD model. However some differences are still observable due to the action of mirror instabilities during the evolution of the plasma. The mirror instability is responsible for changes in the velocity distribution by slowing the gas and reducing the effective sonic Mach number \cite[see also][]{kowal11}. Another feature that can be observed from Figs.  \ref{densk2}-\ref{vm2}  is that the number of structures that develop in KMHD is considerably larger than that obtained in MHD. Our results show that even at late times important imprints of the pressure anisotropy are clearly observed, thus indicating that for a more accurate description of the evolution of this plasma the effects of pressure anisotropy should be included in the simulations.

\begin{figure}[h]
\begin{center}
\includegraphics[width=16pc]{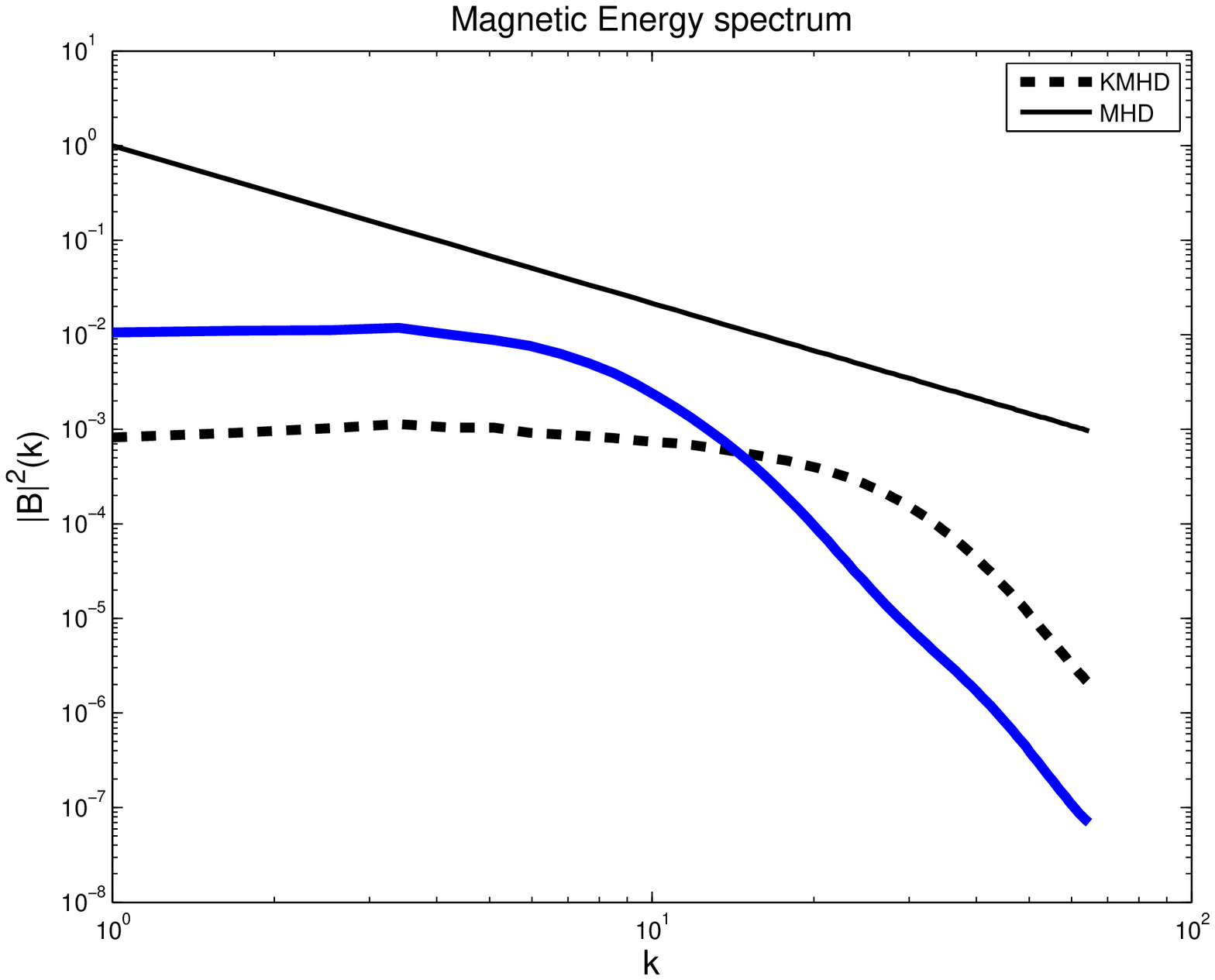}\quad
\includegraphics[width=16pc]{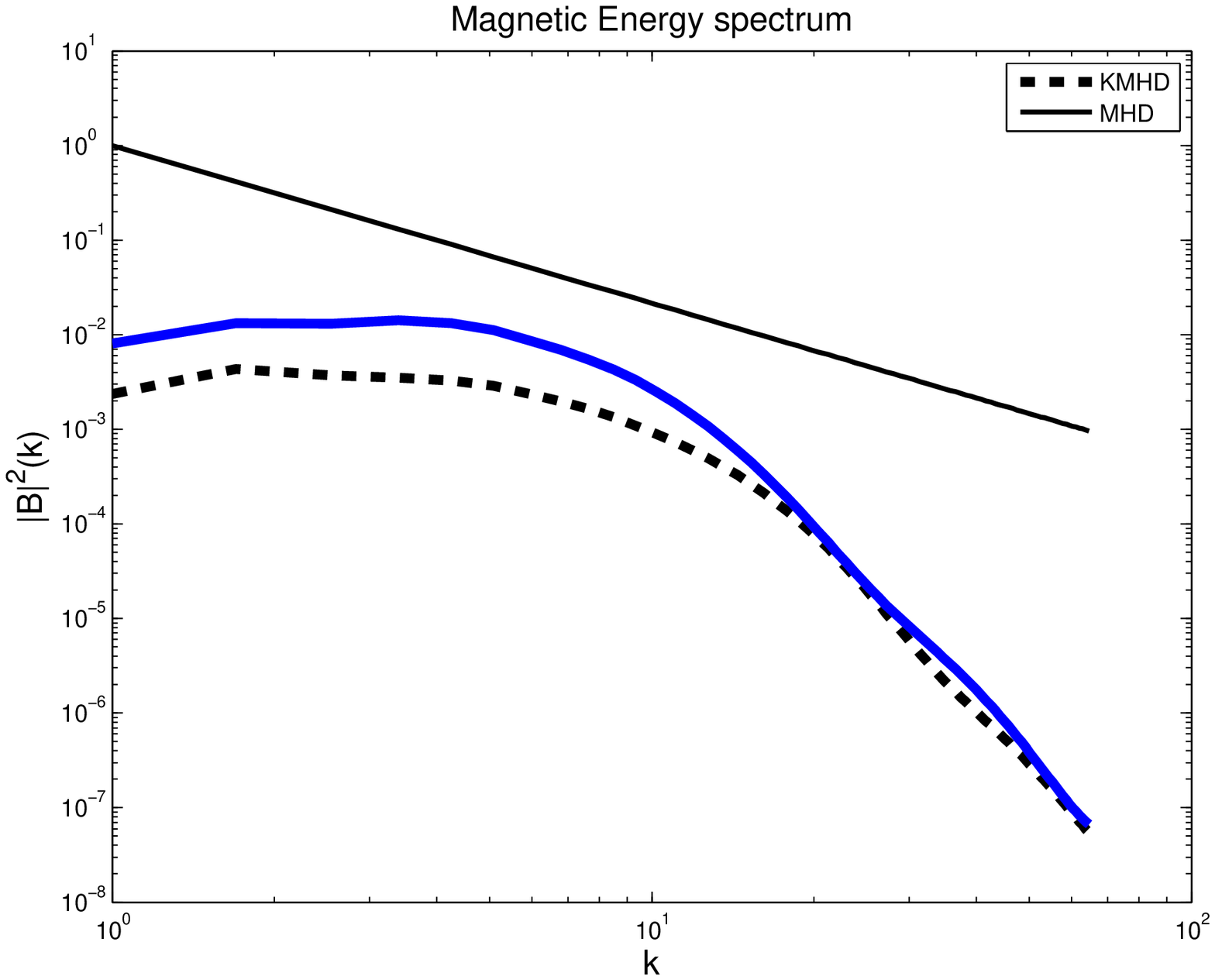}
\end{center}
\caption{\label{esp2}Magnetic energy spectrum corresponding to KMHD (dashed black line) and 
MHD (solid blue line) at t=2 (left) and t=5 (right). The black line corresponds a slope of $-5/3$. }
\end{figure}

The correlation between different scales can be characterized by the energy spectrum,
whose evolution is shown in Fig.\hspace{0.1cm}\ref{esp2} for both formalisms. It is seen that the spectrum in
KMHD evolves towards its MHD counterpart. At late times, the slope of the spectra of both
formalisms is similar. 
Besides providing crucial information
on scale correlations of the plasma, the analysis of the simulated energy spectra is also
important because the slope of the spectra is one of the quantities that can be indirectly
determined by observations using synthetic rotation maps with the observed polarization coming
from radio sources in the ICM \cite{murgia04}. The results of
previous numerical works based on MHD \cite{jones10} seems to favor a power law spectrum with a $k\sim -5/3$ slope, providing support for the use of this value in the 
construction of synthetic rotation maps \citep[see e.g.][]{bonafede10a}. In Fig.\hspace{0.1cm}\ref{esp2} we have added the Kolmogorov slope of $-5/3$ (black line). We note that our results are preliminary and, therefore, further studies are still required to determine the spectrum slope more precisely.   


\vspace{0.4cm}
\section{Conclusions}
\vspace{0.2cm}

As an example of the application of the KMHD with a variable $a_\parallel/ a_\perp$, we compare this formalism with the MHD one 
performing turbulence simulations for astrophysical conditions, namely those prevailing in the ICM, under a Godunov scheme. We find that pressure anisotropy changes have significant impact on the turbulent properties of nearly collisionless magnetized plasmas, such as the ICM. Particularly, the three observables presented here (plasma density, magnetic field and velocity field) show a much more granular structure in the KMHD formalism. The differences can be understood from the action of the mirror instabilities which are fully developed for values of the pressure anisotropies studied here ($a_\parallel/ a_\perp \le 2$)  (see Figs. \ref{densk1}-\ref{vm2}).

The development of  kinetic instabilities, such as these, which arise in the KMHD regime from pressure anisotropy, causes the accumulation of energy at the smaller scales. This may have important consequences on the formation of  large scale structures in the intergalactic medium (IGM) and the ICM, and also in the dynamo amplification of cosmic magnetic fields \cite[e.g.][]{santos-lima11,kowal11,nakwacki11}. A more detailed study of the relation between pressure anisotropy and structure formation in the ICM and IGM is currently in progress \cite{nakwacki11,santoslima12}.

\ack
 This work is partially supported by the Brazilian agencies FAPESP and CNPq.


\bibliographystyle{iopart-num} 
\bibliography{clusbib} 

\providecommand{\newblock}{}
\begin{thebibliography}{10}
\expandafter\ifx\csname url\endcsname\relax
  \def\url#1{{\tt #1}}\fi
\expandafter\ifx\csname urlprefix\endcsname\relax\def\urlprefix{URL }\fi
\providecommand{\eprint}[2][]{\url{#2}}

\bibitem{bonafede10a}
{Bonafede} A, {Feretti} L, {Murgia} M, {Govoni} F, {Giovannini} G and {Vacca} V
  2010 {\em ArXiv e-prints\/} (\textit{Preprint} \eprint{1009.1233})

\bibitem{schekochihin06}
{Schekochihin} A~A and {Cowley} S~C 2006 {\em Physics of Plasmas\/} {\bf 13}
  056501 (\textit{Preprint} \eprint{arXiv:astro-ph/0601246})

\bibitem{Barakat82}
{Barakat} A~R and {Schunk} R~W 1982 {\em Plasma Physics\/} {\bf 24} 389--418

\bibitem{krall}
{Krall} N~A and {Trivelpiece} A~W 1973   {USA: McGraw--Hill}

\bibitem{quest96}
{Quest} K~B and {Shapiro} V~D 1996 {\em \jgr\/} {\bf 1012} 24457--24470

\bibitem{nakwacki11}
{Nakwacki} M~S, {de Gouveia dal Pino} E~M, {Kowal} G and {Santos-Lima} R 2012
  in preparation

\bibitem{bogdanovic10}
{Bogdanovi{\'c}} T, {Reynolds} C~S and {Massey} R 2011 {\em \apj\/} {\bf 731} 7
  (\textit{Preprint} \eprint{1005.2193})

\bibitem{kowal11}
{Kowal} G, {Falceta-Gon{\c c}alves} D~A and {Lazarian} A 2011 {\em New Journal
  of Physics\/} {\bf 13} 053001 (\textit{Preprint} \eprint{1012.5125})

\bibitem{cgl}
{Chew} G~F, {Goldberger} M~L and {Low} F~E 1956 {\em Royal Society of London
  Proceedings Series A\/} {\bf 236} 112--118

\bibitem{kowal10}
{Kowal} G and {Lazarian} A 2010 {\em \apj\/} {\bf 720} 742--756

\bibitem{kowal07}
{Kowal} G, {Lazarian} A and {Beresnyak} A 2007 {\em \apj\/} {\bf 658} 423--445
  (\textit{Preprint} \eprint{arXiv:astro-ph/0608051})

\bibitem{kowal09}
{Kowal} G, {Lazarian} A, {Vishniac} E~T and {Otmianowska-Mazur} K 2009 {\em
  \apj\/} {\bf 700} 63--85 (\textit{Preprint} \eprint{0903.2052})

\bibitem{falceta08}
{Falceta-Gon{\c c}alves} D, {Lazarian} A and {Kowal} G 2008 {\em \apj\/} {\bf
  679} 537--551 (\textit{Preprint} \eprint{0801.0279})

\bibitem{rosin10}
{Rosin} M~S, {Schekochihin} A~A, {Rincon} F and {Cowley} S~C 2011 {\em
  \mnras\/} {\bf 413} 7--38 (\textit{Preprint} \eprint{1002.4017})

\bibitem{Schekochihin05}
{Schekochihin} A, {Cowley} S, {Kulsrud} R, {Hammett} G and {Sharma} P 2005 {\em
  The Magnetized Plasma in Galaxy Evolution\/} ed {K~T~Chyzy,
  K~Otmianowska-Mazur, M~Soida, \& R-J~Dettmar } pp 86--92

\bibitem{howes06}
{Howes} G~G, {Cowley} S~C, {Dorland} W, {Hammett} G~W, {Quataert} E and
  {Schekochihin} A~A 2006 {\em \apj\/} {\bf 651} 590--614 (\textit{Preprint}
  \eprint{arXiv:astro-ph/0511812})

\bibitem{murgia04}
{Murgia} M, {Govoni} F, {Feretti} L, {Giovannini} G, {Dallacasa} D, {Fanti} R,
  {Taylor} G~B and {Dolag} K 2004 {\em \aap\/} {\bf 424} 429--446
  (\textit{Preprint} \eprint{arXiv:astro-ph/0406225})

\bibitem{jones10}
{Jones} T~W, {Porter} D~H, {Ryu} D and {Cho} J 2011 {\em \memsai\/} {\bf 82}
  588 (\textit{Preprint} \eprint{1101.4050})

\bibitem{santos-lima11}
{Santos-Lima} R, {de Gouveia Dal Pino} E~M, {Lazarian} A, {Kowal} G and
  {Falceta-Gon{\c c}alves} D 2011 {\em IAU Symposium\/} ({\em IAU Symposium\/}
  vol 274) pp 482--484 (\textit{Preprint} \eprint{1102.5139})

\bibitem{santoslima12}
{Santos-Lima} R, {de Gouveia dal Pino} E~M, {Kowal}, {Nakwacki} M~S,
  {Falceta-Gon{\c c}alves} D and {Lazarian} A 2012   in preparation

\end{thebibliography}
 {\typeout{}
  \typeout{****************************************************}
  \typeout{****************************************************}
  \typeout{** Please run "bibtex \jobname" to optain}
  \typeout{** the bibliography and then re-run LaTeX}
  \typeout{** twice to fix the references!}
  \typeout{****************************************************}
  \typeout{****************************************************}
  \typeout{}
}
\end{document}